\documentclass[12pt]{article}
\begin{document}
\renewcommand{\thefootnote}{\fnsymbol{footnote}} 
\centerline{\Large\bf Hadronic string and chiral symmetry 
breaking\footnote{Talk given at
12th International Seminar on High Energy Physics QUARKS'2002
                        Novgorod, Russia, June 1-7, 2002}}

\bigskip

\centerline{
\large A. A. Andrianov}
\centerline{St.Petersburg State University, Russia and INFN, Bologna, 
Italy}
\centerline{and}
\centerline{\large D. Espriu}
\centerline{University of Barcelona, Spain}

\bigskip

\centerline{\bf Abstract}
{\small
We assume that QCD can be effectively described
with string-like variables.
The hadronic string is built over the 
chirally non-invariant QCD vacuum by means of the boundary interaction
with background chiral fields associated with pions. By making this interaction
compatible with the conformal symmetry of the string and 
with the unitarity constraint 
on chiral fields we reconstruct the equations of motion for the latter ones 
and furthermore recover the Lagrangian of non-linear sigma model of pion 
interactions. The estimated chiral structural constants of 
Gasser and Leutwyler 
fit well the phenomenological values.}\\

\centerline{\large\bf 1. Introduction}

\medskip

The history of attempts to describe the hadrons in the framework of
a string theory beyond or within QCD encompasses already  
more than 30 years (see,\cite{Ven}-\cite{solo} as well as 
the reviews \cite{Rebbi}-\cite{pol}). 
The commonly cited arguments to justify the stringy description of QCD
 are the dominance of planar gluon diagrams
in the large $N$ limit\cite{largeN} 
being interpreted as the world-sheet of a string,
the expansion in terms of surfaces built out of plaquettes
in strong-coupling lattice QCD\cite{lattice}, and 
the incarnation of
Regge phenomenology\cite{regge} within QCD\cite{Lip}.

There is a motivated 
agreement that in a certain
kinematic regime the  Nambu-Goto or the Polyakov
string action may be satisfactory. Here we focus on 
low-energy properties of string-generated particle states 
and it is known for a long
time that the hadronic amplitudes derived from such type of strings
are not quite physically consistent.  
To illuminate their flaws we recall
the original Veneziano amplitude\cite{Ven}, which can be derived from
Nambu-Goto string and supposedly describes the 
scattering amplitude of four
pions. One can show that in this amplitude the scalar resonance is  
a tachyon and the vector state (which we should
identify with the rho particle) is massless. 
At last such an amplitude does not have the appropriate Adler
zero, i.e. the property that at $s=t=0$ the pion scattering amplitude 
vanishes.

It is quite conceivable that the main reason for the
presence of a tachyon in the spectrum and the wrong chiral
properties lies in a wrong choice
of the vacuum\cite{tachyon}.
A possible way to take into account the non-perturbative properties
 of the QCD vacuum 
was suggested in \cite{ADE} and developed in \cite{aabe}. Namely, one can assume
that in QCD chiral symmetry breaking takes place and
the massless (in the chiral limit) pseudoscalar mesons form the background
of the QCD vacuum,
whereas other massive excitations are assembled into a string.
The massless pion fields can be collected in a
unitary matrix $U(x)$ belonging to $SU(2)$ group (here we 
consider non-strange Goldstone mesons only). It describes excitations
around the non-perturbative vacuum breaking the chiral symmetry. From the 
string point of view
$U(x)$ is nothing but a
bunch of couplings involving the string variable $x_\mu(\tau,\sigma)$. 
It has to be coupled to the
boundary of the string where flavor is attached.
Our goal is to find
a consistent string propagation in this non-perturbative background.

An essential property of string theory is conformal invariance. 
Since it must hold when 
perturbing the string around any vacuum we demand the new coupling to
chiral fields, living on the boundary, to preserve it.

Thus our proposal is to introduce the general reparameterization-invariant
boundary interaction to chiral fields and derive all the divergences induced by
this interaction. We shall need additional dimensional operators
 in the boundary action to renormalize divergences. 
From the condition of vanishing $\beta$ functions for $U(x)$
the equations of motion for chiral fields are obtained in the low-momentum 
(derivative)
expansion. We consistently implement the unitarity constraint on the 
chiral fields 
and locality of the chiral
Lagrangian and finally calculate the $O(p^4)$ terms of 
the Gasser and Leutwyler\cite{GL} 
effective Lagrangian. A strikingly good correspondence
with their phenomenological values is found.\\

\centerline{\large\bf 2. Pion interaction to the QCD string and Diagrammar}

\medskip

The hadronic string in the conformal gauge is described by the following
conformal field theory action which has four dimensional Euclidean
space-time as target space
\begin{equation}
{\cal W}_{str}=\frac{1}{4\pi\alpha'}\int d^{2+\epsilon}\sigma
\left(\frac{\varphi}{\mu}\right)^{-\epsilon} 
\partial_i x_\mu \partial_i x_\mu, 
\label{string}
\end{equation} 
where for $\epsilon = 0$ one takes 
$x_\mu = x_\mu(\tau, \sigma)$, $ 
-\infty <\tau< \infty, 0< \sigma <\infty$,
$i = \tau,\sigma$, $\mu=1,...,4$. 
The conformal factor $\varphi(\tau, \sigma)$ is 
introduced to restore the conformal 
invariance in $2+\epsilon$ dimensions. The Regge trajectory slope
(related to the inverse string tension) is known
to be universal $\alpha' \simeq 0.9$ GeV$^{-2}$ \cite{lia}.

We would like to couple in a chiral invariant manner
the matrix in flavor space $U(x)$ containing  the meson fields to the
string degrees of freedom while preserving
general covariance in the two dimensional coordinates and conformal
invariance under local scale transformations
of the two-dimensional metric tensor.

Since the string variable $x$ does not contain any flavor dependence,
we introduce two dimensionless Grassmann variables (`quarks')
living on the
boundary of the string sheet: 
$\psi_L(\tau),\psi_R(\tau)$. They transform in the fundamental representation
of the light flavor group ($SU(2)$ in the present paper).
A local hermitean action $S_b = \int d\tau L_f$ is then introduced 
on the boundary
$ \sigma =0$ to describe the interaction with background chiral fields 
$U(x(\tau)) = \exp(i \pi(x)/f_\pi)$, where the normalization scale is
set to $f_\pi\simeq 93 MeV$, 
the weak pion decay constant.

The boundary Lagrangian is chosen to be reparameterization invariant
and in its minimal form reads 
\begin{eqnarray}
L_f&=&\frac12 i \left(\bar\psi_L U (1 - z) \dot\psi_R  - 
\dot{\bar\psi_L} U (1 +z)\psi_R \right.\nonumber\\
&&\left.+ \bar\psi_R U^+ (1 + z^*)
\dot\psi_L - \dot{\bar\psi_R} U^+ (1 - z^*) \psi_L\right), \label{lag}
\end{eqnarray}
herein and further on a dot implies a $\tau$ derivative: 
$\dot\psi \equiv d\psi/d\tau$. 

A further restriction is obtained by requiring $CP$ invariance,
\begin{equation}
U \leftrightarrow U^+, \quad \psi_L  \leftrightarrow  \psi_R.
\label{CP1}
\end{equation}
The above Lagrangian is $CP$ symmetric for $ z = - z^* = ia$.  
The fulfillment of this symmetry happens to be crucial to
preserve conformal symmetry in the presence of the added boundary interaction.

Now we expand the function $U(x)$ in powers of the string coordinate 
field $x_\mu(\tau) =x_{0\mu} + \tilde x_\mu(\tau) $ around a constant $x_0$,
\begin{equation}
U(x) = U(x_0) + \tilde x_\mu(\tau) \partial_\mu U(x_0) + 
\frac12 \tilde x_\mu(\tau) \tilde x_\nu(\tau) \partial_\mu\partial_\nu U(x_0) 
+\ldots. 
\label{expan}
\end{equation}
and look
for the potentially divergent
one particle irreducible diagrams.The two-fermion, 
$N$-boson vertex operators are generated by the expansion 
(\ref{expan}), from the generating functional 
$ Z_b = \langle\exp(i S_b)\rangle$ and eq.(\ref{lag}).
Each additional loop comes with a power of $\alpha^\prime$.
One can find a resemblance to the familiar 
derivative expansion of chiral perturbation theory \cite{GL}.

The free fermion propagator is
\begin{equation}
\langle\psi_R (\tau) \bar\psi_L(\tau')\rangle =
\langle\psi_L (\tau) \bar\psi_R(\tau')\rangle^\dagger = U^{-1} (x_0) 
\theta(\tau - \tau'),
\end{equation}
if we impose $CP$ symmetry for unitary chiral fields $U(x)$.

The free boson propagator projected on the boundary is
\begin{equation}
\langle x_\mu(\tau) x_\nu(\tau')\rangle = \delta_{\mu\nu}
\Delta (\tau -\tau') =- 2 \delta_{\mu\nu} \alpha'\ln(|\tau 
-\tau'|\mu).
\end{equation}
The normalization of the string propagator is inferred \cite{aabe} from 
the  
definition of the kernel of the N-point tachyon amplitude for the 
open string \cite{Rebbi}. In dimensional regularization one adopts
$\Delta (0) \sim
\alpha'/\epsilon$ and  $\Delta'(0) = 0$. 

To implement the
renormalization process we perform a loop (equivalent to a
derivative) expansion, proceed to determine the counterterms required
to make the theory finite and further on to impose a vanishing
beta functional for the coupling $U(x)$ to implement
the absence of conformal anomaly.\\

\centerline{\large\bf 3. Renormalization at one and two loops}

\medskip

Using the above set of Feynman rules one
arrives at the one-loop divergent part of the propagator,
\begin{equation}
- \theta(A - B) U^{-1} \delta U U^{-1},\quad \delta U \equiv \Delta(0) 
\left[\frac12 \partial^2_{\mu} U -
\frac{3 + z^2}{4}\partial_\mu U U^{-1}\partial_\mu U\right] .\label{1div}
\end{equation}
This divergence is eliminated by introducing 
an appropriate counterterm $U \to U+\delta U$.
Conformal symmetry is restored (the beta-function is zero) if
the above contribution vanishes, $\delta U = 0$.

Let us find out for which value of $z$ this variation of $U$ is 
compatible with 
its unitarity.
\begin{equation}
\delta (U U^+)= U \cdot\delta U^+ + \delta U\cdot U^+ = 0. \label{unit}
\end{equation}
A simple calculation shows that this
takes place for $z = \pm i$. 
The related local classical action which 
has $\delta U = 0$ as equation of motion is
\begin{equation}
W^{(2)} = \frac{f_\pi^2}{4}\int d^4 x \mbox{\rm tr}\left[\partial_\mu U
\partial_\mu U^+ \right], \label{weinb}
\end{equation}
i.e. the well known non-linear 
sigma model of  pion interactions.

We have thus found the chiral action induced by the QCD string.
It has all the required properties of locality, chiral symmetry 
and proper low momentum behavior (Adler zero) and describes
massless pions. However $f_\pi$, the overall normalization scale, cannot be
predicted from these arguments. 

Before proceeding to a full two loop calculation we have to check whether 
the minimal
Lagrangian (\ref{lag}) is sufficient to renormalize also the vertices containing
the boson legs.  It turns out that it is not.

To obtain the divergences for vertices with external boson lines
we introduce an external background boson field $\bar x_\mu $ 
and split $ x_\mu =  \bar x_\mu + \eta_\mu $. 
The free propagator 
for the fluctuating field $\eta_\mu $ coincides with the one for  $x_\mu $.

The  total one-loop divergence in the vertex with two fermions and 
one boson line can be represented by the following
vertex 
operator in the Lagrangian
\begin{equation}
\frac{i}{2}\left( \bar\psi_L\Phi^{(1)}\dot\psi_R -\dot{\bar\psi_L}\Phi^{(2)}
\psi_R\right)+ \mbox{\rm h.c.},\quad 
\Phi^{(1,2)} \equiv \bar x_\mu(\tau)(1\mp z)\left[  
\partial_{\mu}\left(\delta U\right)\mp\phi_\mu\right].\label{adddiv}
\end{equation}
The terms proportional to derivatives of $\delta U$ 
are automatically eliminated by the renormalization of
 the one-loop propagator. 
But the part proportional to $\phi _\mu$ remains and to absorb 
these divergences new counterterms are required. The latter ones 
can be parameterized with three bare 
constants $g_1$ , $g_2$ and $g_3$, which are real 
if the $CP$ symmetry for $z = - z^*$ holds
\begin{eqnarray}
\Delta L_{bare}&= &\frac{i}{8} (1 - z^2)\bar\psi_L\left( 
(g_1 - z g_2) \partial_\nu \dot{U} U^{-1}\partial_\nu U 
- (g_1 + z g_2) \partial_\nu U U^{-1}\partial_\nu \dot{U} \right.\nonumber\\
&&\left.+  2z g_3 \partial_\nu U U^{-1} \dot{U} U^{-1} \partial_\nu U 
\right)\psi_R  + \mbox{\rm h.c.} \label{count}
\end{eqnarray} 
Renormalization is accomplished by subtraction,
\begin{equation}
g_i = g_{i,r} - \Delta(0). \label{gren}
\end{equation}
The constants $g_{i,r}$ are finite, but in principle scheme dependent.
The counterterms are of  higher dimensionality than 
the original Lagrangian (\ref{lag}) and the couplings $g_i$ are 
of dimension $\alpha'$. Since (\ref{lag}) was the most 
general coupling permitted by the symmetries of the model, one concludes that
conformal symmetry seems to be broken by these boundary couplings 
already at tree level. 
However in spite of the fact that 
the new couplings are dimensional, it turns out \cite{aabe} that 
their contribution into the trace of the energy-momentum tensor 
vanishes once the requirements of unitarity of $U$ and $CP$ invariance are 
taken into account. Therefore conformal invariance is not broken 
at the order
we are working. 

On the other hand
the appearance of new vertices changes the  
fermion propagator. 
One obtains from such terms the following contribution to the propagator 
\begin{eqnarray}
&&\theta(A-B)\frac{1}{16}\Delta(0) (1-z^2)
U^{-1} \left\{2 (g_{1,r} - z^2 g_{2,r})
 \partial_{\rho} U  U^{-1} \partial_\mu\partial_\rho U  U^{-1}\partial_{\mu} U\right.\nonumber\\ 
&&\left.- (1+z) (g_{1,r} + z g_{2,r}) \partial_{\rho} U  U^{-1} \partial_\mu U  U^{-1} \partial_\rho\partial_{\mu} U\right.\nonumber\\ 
&&\left.
- (1-z) (g_{1,r} - z g_{2,r}) \partial_{\rho}\partial_\mu U  U^{-1} \partial_\rho U  U^{-1}\partial_{\mu} U\right.\nonumber\\ 
&&\left.
+4z^2 g_{3,r} \partial_{\rho} U  U^{-1} \partial_\mu U  U^{-1} \partial_\rho U  U^{-1}\partial_{\mu} U
\right\} U^{-1} \nonumber\\
&\equiv& - \theta(A-B)\Delta(0) U^{-1}  \delta^{(4)}U  U^{-1}, 
\label{dren}
\end{eqnarray}
One should  add 
this divergence to the one-loop result, thereby 
modifying the 
$U$ field renormalization and equations of motion
\begin{equation}
\bar\delta U = \Delta(0) \left[\frac12 \partial^2_{\mu} U - 
\frac{3 + z^2}{4}\partial_\mu U U^{-1}\partial_\mu U + \delta^{(4)}U\right]= 0.
\label{eom4}
\end{equation}
This is one source of $O(p^4)$ terms and we shall
see that there is another contribution at two loops.

As to other vertices it can be proved \cite{aabe} that
 any diagram with an arbitrary number of external
boson lines and two fermion lines, i.e. any vertex of those generated by the
perturbative expansion of
(\ref{lag}) is rendered finite by the previous counterterms. This completes
the renormalization program at one loop.

There are 10 two-loop one-particle irreducible diagrams which are 
analytically calculated 
in \cite{aabe}.
The divergences in the propagator consist of the double divergent
part, 
$ \sim \Delta^2(0)$ and of
the single divergent contributions,  
$\sim \Delta (0)$. The substantial part of these divergences is fully 
renormalized by performing the one-loop renormalization and taking into
account the renormgroup evolution.

Some single-pole divergences remain however. Namely, there are 
 divergences linear in
 $\Delta(0)$ which come from irreducible two-loop
diagrams with maximal number of vertices.
These  divergences appear to be\\

\begin{eqnarray}
- \Delta(0) U^{-1} \delta^{(4)}_{2-l} U^{-1} &\equiv&  c \Delta(0)\left[ 
U^{-1}
 \partial_{\rho} U  U^{-1} \partial_\mu U  
U^{-1} \partial_\mu U  U^{-1}\partial_{\rho} U  U^{-1} -\right.\nonumber\\
&&\left.- U^{-1} 
 \partial_{\rho} U  U^{-1} \partial_\mu U  U^{-1} \partial_\rho U  U^{-1}\partial_{\mu} U  U^{-1}\right].
\label{dfive}
\end{eqnarray}
with $c = \alpha'(1-z^2)^2/8 = \alpha'/2$ for $z = \pm i$.
This term survives after adding all the counterterms.
It must therefore modify
the equation of motion (ref{eom4}) at the next order in the $\alpha^\prime$ expansion,
$\delta^{(4)}U \rightarrow 
\delta^{(4)} U +  \delta^{(4)}_{2-l}$. Its presence
allows for non zero solutions for the coupling constants 
$g_i$ and therefore for nonzero values for the
Gasser-Leutwyler $O(p^4)$ coefficients.\\

\centerline{\large\bf 4. Local integrability of equations of motion}

\medskip 

The equation of motion, $\delta U = 0$, 
can be obtained from the dimension-two local action 
(\ref{weinb}), 
involving 
a unitary matrix $U(x)$, only for $z=\pm i$.
If the four-derivative part of equations of motion 
can be derived from dimension-four operators in a local effective 
Lagrangian
then certain constraints are to be imposed on constants $g_{i,r}$.

Such a Lagrangian has only two terms compatible with 
the chiral symmetry,
\begin{equation}
{\cal L}^{(4)} = f_\pi^2 \mbox{tr}\left( K_1 \partial_\mu U \partial_\rho  
U^+ \partial_{\mu} U \partial_{\rho} U^+ 
+ K_2  \partial_\mu U \partial_{\mu} U^+ \partial_{\rho} U  
\partial_{\rho} U^+ \right). \label{dim4}
\end{equation}
Other terms containing $\partial_\mu^2 U$ are reduced to the set (\ref{dim4}) 
with the help  
of  the dimension-two equations of motion.

Variation of the previous Lagrangian supposedly saturate the 
dimension-four component of the equations of motion. Therefrom we identify 
this parameterization of constants with the coupling
constants arising from the equations of motion (\ref{eom4}) supplemented 
with (\ref{dfive}) and after applying the $O(p^2)$ equations of motion. 
Then
one obtains the following set of coefficients for the various chiral 
field structures
\begin{eqnarray}
-2(2K_1 + K_2) &=& \frac{1}{16} (1-z^2)(1\pm z)(g_{1,r} \pm z g_{2,r});
\nonumber\\
-4K_2 &=& \frac18 (1-z^2)(- g_{1,r} + z^2 g_{2,r});\quad 
2[(1-z^2) K_1 + K_2] = -c;\nonumber\\
 -2z^2 K_2&=& 0;\quad 
4[K_1 + K_2] = - \frac14 (1-z^2)z^2g_{3,r} + c,
\end{eqnarray}
For  $z^2 = -1$ only  one solution is possible, 
\begin{equation}
K_2 = 0, \quad K_1 = -\frac14 c = - \frac{\alpha' }{8};\quad\quad
g_{1,r} = - g_{2,r} =  - g_{3,r} = 4 c.
\end{equation}
Thus, comparing eq.(\ref{dim4}) with the usual parameterization
of the Gasser and Leutwyler Lagrangian\cite{GL},   
\begin{equation}
L_1 = \frac12 L_2 = -\frac14 L_3 = - \frac12 K_1 f_\pi^2 = 
\frac{ f_\pi^2\alpha' }{16}.
\end{equation}
For $\alpha' = 0.9$ GeV$^{-2}$ and $f_\pi \simeq 93$ MeV it 
yields $L_2 \simeq 0.9 \cdot 10^{-3}$ 
which is quite a satisfactory result\cite{expt}.

The relation
$L_1 = 1/2 L_2 = -1/4 L_3$ was
established earlier
in bosonization models \cite{AnBo}
and in the chiral quark model\cite{Espr} by means of
a derivative expansion of quark determinant. However at that time its possible
connection with a string
description of QCD was not recognized. The first attempt to derive the
chiral coefficients from the Veneziano-type dual amplitude was
undertaken in\cite{PV} where  a similar relation was found but with different numerical values for the $L_i$. 
However the specific choice of dual amplitude in \cite{PV}
cannot be related to any known hadron string.

Another check
comes from the compatibility of the unitarity of $U$ 
and the equations of motion at the two-loop 
level. It turns out that if 
one accepts arbitrary real coefficients
in the set of dimension-four operators then the 
only solution
compatible with the unitarity is given by the parameterization with 
constants $K_1$ and
$K_2$. \\

In our talk we have reported on a simplified model of the QCD string.
Requiring of its conformal invariance 
around a chirally non-invariant vacuum leads to the Gasser and
Leutwyler
Lagrangian. However the  bosonic string
action used here does not prevent large Euclidean 
world sheets from crumpling \cite{pol1}. It does not also describe correctly
the high-temperature behavior of large $N$ QCD \cite{polch2}. To correct it,
a QCD induced string must be modified \cite{pol1,diam}  including 
operators breaking manifestly conformal symmetry
on the world-sheet for large strings. Nevertheless we are concerned here
with the low-energy string properties and therefore do not expect that the strategy
and technique to derive the chiral field action needs any significant
changes to be adjusted to a modified QCD string action. 

We have restricted ourselves here to the $SU(2)$ 
global flavor
group. In this case only parity-even terms in the equations 
of motion can be
revealed from the simple fermion Lagrangian (\ref{lag}) and to obtain the parity-odd
WZW Lagrangian relevant for the case of three flavors one has to extend the
boundary fermion action supplementing one-dimensional fermions 
with true spinor degrees of freedom.\\

We express our gratitude to the organizers of the
International Workshop QUARKS 2002 in Novgorod for
hospitality.
This talk is supported by
MCyT FPA2001-3598 and CIRIT 2001SGR-00065 and the European
Networks EURODAPHNE and EUROGRID, as well as by Grant RFFI
01-02-17152, Russian
Ministry of Education Grant E00-33-208 and by The Program {\sl Universities
of Russia: Fundamental Investigations} (Grant 992612). By this work A.A. and 
D.E. contribute 
to the fulfillment of  Project INTAS 2000-587.


\begin{thebibliography}{99}
\bibitem{Ven} G. Veneziano, Nuovo Cim. 57A (1968) 190.
\bibitem{lovelace} C. Lovelace, Phys. Lett. 28B (1968) 264;\\
J. Shapiro, Phys. Rev. 179 (1969) 1345.
\bibitem{NG} Y. Nambu, in {\it ``Symmetries and Quark Models"},
R. Chand, ed., Gordon and Breach, 1970;\\ 
L. Susskind, Nuovo Cim. 69A (1970) 457.
\bibitem{rigid} A.M. Polyakov, Phys. Lett. 103B (1981) 207.
\bibitem{pronko} G.P. Pronko and A.V. Razumov, Theor. Math. Phys. 56 
(1984) 760.
\bibitem{klein} H. Kleinert,
Phys. Lett. 174B (1986) 335.
\bibitem{ant} D. Antonov, D. Ebert, and Y.A. Simonov, 
Mod. Phys. Lett. A11 (1996) 1905.
\bibitem{solo} L.D. Solovev, Phys. Rev. D58 (1998) 035005.
\bibitem{Rebbi} C.Rebbi,
Phys.Rep.C12 (1974) 1.
\bibitem{regge} P. Frampton, {\it ``Dual Resonance
Models"}, Benjamin, 1974.
\bibitem{pol} A.M.Polyakov,  {\it ``Gauge Fields and Strings''}, Harwood,
Chur, Switzerland, 1987.
\bibitem{largeN} G. 'tHooft, Nucl. Phys. B72 (1974) 461; \\
G. Veneziano, Nucl. Phys. B117 (1976) 519.
\bibitem{lattice} K.G. Wilson, Phys. Rev. D10 (1974) 2445.
\bibitem{Lip} R. Kirschner, L.N. Lipatov and L. Szymanowski, 
Nucl. Phys. B425 (1994) 579.
\bibitem{tachyon} E. Cremmer and J. Scherk, Nucl. Phys. B72 (1974) 117.
\bibitem{ADE} J. Alfaro, A. Dobado and D. Espriu,
 Phys. Lett. B460 (1999) 447.
\bibitem{aabe} J.Alfaro, A.A.Andrianov, L.Balart and D.Espriu,
{\it ``Hadronic string, conformal invariance and chiral symmetry''},
hep-th/0203215 (2002) 35p.
\bibitem{GL} J. Gasser and H. Leutwyler, Nucl. Phys. B250 (1985) 465.
\bibitem{lia} S.  Filipponi, G. Pancheri and  Y. Srivastava,
Phys. Rev. Lett. 80 (1998) 1838.
\bibitem{expt} G. Amoros, J. Bijnens, P. Talavera,
Nucl. Phys. B602 (2001) 87. 
\bibitem{AnBo} A.A. Andrianov and L.Bonora, Nucl. Phys. B233 (1984) 232; 247\\
 A.A. Andrianov, Phys.Lett.B157 (1985) 425. 
\bibitem{Espr}D. Espriu, E. de Rafael and J. Taron, Nucl. Phys. B345 (1990) 22:
 Erratum-{\it ibid.} B355 (1991) 278.
\bibitem{PV} M. Polyakov and V. Vereshagin, Phys. Rev. D54 (1996) 1112.
\bibitem{pol1} A.M. Polyakov, Physica Scripta T15 (1987) 191.
\bibitem{polch2}J. Polchinski and Zhu Yang, Phys. Rev. D46 (1992) 3667.
\bibitem{diam}M.C. Diamantini, H. Kleinert and C.A. Trugenberger, Phys. Rev.
Lett. 82 (1999) 267.
\end{thebibliography}
\end{document}